\documentclass[11pt]{article}
\usepackage[margin=0.95in]{geometry}
\usepackage{epsfig,ulem,amssymb,soul}

\usepackage{graphicx} 
\usepackage{epstopdf}
\usepackage{amsfonts,amsmath}
\usepackage{bm}
\usepackage{setspace}
\usepackage{comment}
\usepackage{siunitx}
\usepackage{cancel}
\usepackage{mathrsfs}
\usepackage{multirow}
\usepackage{subcaption}
\usepackage[hang,flushmargin,bottom]{footmisc}
\usepackage{mathtools}
\usepackage{lipsum}
\usepackage{array}
\usepackage{enumerate}
\usepackage{ifthen}
\usepackage[affil-it]{authblk}
\usepackage{cite}
\usepackage{textcomp}
\usepackage{leftidx}
\usepackage{relsize}
\usepackage{braket}
\usepackage{float}

\usepackage[autostyle]{csquotes}
\usepackage[dvipsnames]{xcolor}
\usepackage[breaklinks=true,colorlinks=true,hypertexnames=false,linktoc=section,pdfencoding=auto,psdextra=true,linkcolor=blue,anchorcolor=black,citecolor=blue,filecolor=black,menucolor=black,runcolor=black,urlcolor=black]{hyperref} 
\newcommand{\etal}{\textit{et al}.~}

\providecommand{\keywords}[1]{\textbf{\textit{Keywords---}} #1}

\title{GRIDS-Net: Inverse shape design and identification of scatterers via geometric regularization and physics-embedded deep learning}
\date{}

\author[1]{Siddharth Nair}
\author[2]{Timothy F. Walsh}
\author[2]{Greg Pickrell}
\author{Fabio Semperlotti\thanks{To whom correspondence should be addressed. Email: fsemperl@purdue.edu }}
\affil[1]{Ray W. Herrick Laboratories, School of Mechanical Engineering,  Purdue University, West Lafayette, IN 47907, USA}
\affil[2]{Sandia National Laboratory, Albuquerque, NM 87185, USA}
\setstcolor{red}

\usepackage{lineno}

\begin{document}
\maketitle

\begin{abstract}

This study presents a deep learning based methodology for both remote sensing and design of acoustic scatterers. The ability to determine the shape of a scatterer, either in the context of material design or sensing, plays a critical role in many practical engineering problems. This class of inverse problems is extremely challenging due to their high-dimensional, nonlinear, and ill-posed nature. To overcome these technical hurdles, we introduce a geometric regularization approach for deep neural networks (DNN) based on non-uniform rational B-splines (NURBS) and capable of predicting complex 2D scatterer geometries in a parsimonious dimensional representation. Then, this geometric regularization is combined with physics-embedded learning and integrated within a robust convolutional autoencoder (CAE) architecture to accurately predict the shape of 2D scatterers in the context of identification and inverse design problems. 
An extensive numerical study is presented in order to showcase the remarkable ability of this approach to handle complex scatterer geometries while generating physically-consistent acoustic fields. The study also assesses and contrasts the role played by the (weakly) embedded physics in the convergence of the DNN predictions to a physically consistent inverse design. 

\noindent\keywords{Inverse scattering, Deep learning, Remote sensing, Material design, Auto encoders}
\end{abstract}


\section{Introduction}
\label{ssec: Introduction}

The ability to determine the optimal geometric and material configurations necessary to achieve target functionalities is a typical inverse design problem critical for a wide range of important scientific and engineering applications, including, to name a few, remote sensing \cite{wang2011optimization, entekhabi1994solving}, material design and discovery \cite{sigmund1996some,kim2018deep, noh2019inverse}, nondestructive testing \cite{santamarina1998introduction, altpeter2002robust}, security screening \cite{harding2012x}, and biomedical imaging \cite{macleod1998recent, bertero2006inverse, yaman2013survey}. These applications involve a variety of engineering disciplines, including acoustics \cite{wu2022physics}, photonics \cite{pestourie2018inverse, bayati2020inverse}, and solid mechanics \cite{maute2013topology}. 
The solution to inverse design problems is especially challenging due to the high dimensionality of the design space, which makes exhaustive searches very computationally intensive and the overall problem severely ill-posed. 
With the advent of geometric parameterization techniques \cite{samareh2001survey}, the ability to integrate discrete geometric parameters within traditional inverse design approaches enabled a more parsimonious dimensional representation of the design space and opened opportunities for a more effective search of the design space. Explicit geometry parameterization-based inverse design techniques, typically referred to as \textit{shape optimization} \cite{feijoo2003application, pestourie2018inverse}, and their application to the design of materials for acoustic wave manipulation is the focus of the current study. Note that shape optimization differs from topology optimization, which implicitly parameterizes the inverse design space via density functions or level sets \cite{duhring2008acoustic}.

Early approaches to shape optimization for targeted functionalities combined human intuition with heuristics and trial-and-error approaches \cite{koenderink2015nanophotonics}. These time-consuming and computationally expensive techniques were then replaced or, at least, complemented by numerical iterative minimization schemes in order to eliminate the dependence on pure intuition \cite{molesky2018inverse,pestourie2018inverse}. Despite several recent efforts to develop more powerful shape optimization approaches, and regardless of the specific type of numerical solution technique, the nonlinear behavior and ill-posed nature of the inverse design problem remain formidable challenges to overcome when addressing high dimensional systems \cite{colton1998inverse, habashy2004general}. The inherent nonlinear nature of the inverse design problem requires numerical solvers to deal with non-convex objective functions. Global optimization schemes, such as genetic algorithms \cite{back1997evolutionary, molesky2018inverse}, were developed specifically for this class of problems and allowed converging to the optimal solution without using local search methods (such as first-order gradient descent methods \cite{fu2005simulation, molesky2018inverse}). One of the major drawbacks of this class of methods is the significant computational cost associated with the search process. Popular strategies, like the Tikhonov method \cite{zhong2019multiresolution} and the total variation method \cite{abubaker2001total,zhong2019multiresolution} relied on setting suitable geometric constraints on the inverse design in order to possibly transform it into a well-posed problem and make their solution amenable to the use of iterative global minimization schemes. Although these numerical solvers can address inverse design problems in a constraint design space, they require iterations over a large parameter space whose computational time drastically scales up with the complexity of the design.

In recent years, deep learning techniques have been widely used in many areas of inverse design, including, but not limited to, acoustics \cite{wu2022physics}, photonics \cite{so2020deep, jiang2021deep}, and mechanical materials \cite{guo2021artificial}. In many of the early applications, deep neural networks 
(DNNs) were employed to perform either supervised \cite{tao2019application, guo2021artificial, pestourie2020active, zhang2021multi, white2019multiscale} or unsupervised \cite{liu2018generative, so2019designing} learning to extract the nonlinear mapping between the inverse design and the target functionalities. More specifically, DNNs have been widely used as surrogate models of numerical solvers to accelerate the constrained optimization process. Other studies have focused on developing DNN surrogate models based on the paradigm of reinforcement learning \cite{so2020deep}. These approaches have shown a reasonable degree of success in constructing accurate models that aid the numerical solvers by reducing the computational burden and by improving the shape optimization efficiency in high dimensional scenarios. However, these DNN surrogate model-based inverse design approaches still depend on the use of traditional numerical solvers, hence inheriting the same limitations.

More recently, Raissi $\etal$ \cite{raissi2019physics} introduced the concept of physics-informed neural networks as an alternative method for the numerical solution of partial differential equations (PDEs). Although PINNs have been used to solve inverse problems to evaluate unknown data from partial observations in optics \cite{chen2020physics, lu2021physics}, biomedicine \cite{kissas2020machine}, and fluid mechanics \cite{raissi2020hidden, lu2021physics}, a PINN for inverse design was recently developed by Lu $\etal$ \cite{lu2021physics}. The investigation focused on topology optimization with density function-based geometry parameterization. Although PINN's implementation can be easier than the conventional numerical solvers, an equation (usually a PDE) relating the inverse design (implicitly parameterized by density function or level-set) to the target functionality is an essential requirement for this approach. This requirement is fulfilled in topology optimization, whereas obtaining a mathematical relation connecting the discrete geometry parameters to the target functionality in shape optimization is a complex problem. The above discussion suggests that there is still a significant need for efficient and robust shape optimization techniques and that data-driven DNNs could serve as a foundational framework to overcome several existing limitations. 

Over the years, thanks to their effective yet simple architecture, convolutional neural networks (CNNs) have emerged as the primary data-driven neural network type to solve challenging image-based tasks. CNNs have an excellent ability to learn visual features, which made them very well suited to solve forward and inverse imaging problems \cite{krizhevsky2017imagenet, mccann2017review}. Moreover, the capability to represent target fields as images makes CNNs a potent approach to solve inverse design problems via image-driven pattern recognition. Recently, CNNs have been employed as data-driven deep learning based inverse design approaches in various scientific fields, including photonics \cite{so2020deep}, aerodynamics \cite{sekar2019inverse}, and acoustics \cite{gao2021inverse}. The advent of powerful computational resources (like GPUs) and the ability to learn highly nonlinear functions with relative ease makes CNN-based deep learning an effective tool to solve inverse design problems. Although the use of interconnected nonlinear activation functions in the network architecture intrinsically enables deep networks to learn complex nonlinear processes, the ill-posedness of the inverse problem in a high-dimensional design space still ends up limiting the prediction accuracy in simple CNNs.

Solving a high-dimensional ill-posed inverse design problem is a fundamental mathematical challenge. Studies have investigated the combination of reduced-order modeling with constrained regularization to obtain an efficient deep learning based inverse design approach. Recent investigations have highlighted autoencoder-based DNNs as an ideal architecture to achieve reduced-order models. More specifically, the latent space representation of the autoencoder-based neural network showcases a form of nonlinear dimensionality reduction without losing essential information; this process is somewhat analogous to principal component analysis (PCA) on linear data. Lately, studies have successfully developed \textit{convolutional autoencoders (CAEs)} for efficient dimensionality reduction in nonlinear PDE solvers \cite{maulik2021reduced}, wireless communications \cite{chen2018autoencoder}, and seismic engineering \cite{mousavi2019unsupervised}. More recently, Wu $\etal$ \cite{wu2022physics} developed a classification-based CAE for inverse acoustic scatterer design. Here, the approach could classify a target pressure field to match a predefined set of shapes. Alternatively, Yeung $\etal$ \cite{yeung2021global} introduced a generative adversarial network (GAN) based classification approach to predict material property and scatterer geometry for applications in the field of optics. However, reduced order modeling with modified CAE architectures alone has also been insufficient for accurate inverse design predictions due to the need for high-dimensional data to represent the inverse designs.

Geometry parameterization using the B\'ezier curve, B-spline, and non-rational B-spline (NURBS) introduces a parsimonious dimensional representation of high dimensional geometry. These parameterization methods can help develop an efficient lower-dimensional nonlinear mapping between the target field and the geometry. Among the existing geometry parameterization techniques, NURBS can best represent any basic or free-form shape. This ability to represent a large variety of shapes via a relatively small number of parameters makes NURBS an effective tool for inverse shape design \cite{ma1998nurbs, saini2017nurbs}. Traditionally, NURBS has been an integral part of finite element (FE) based inverse design approaches \cite{hughes2005isogeometric}. In recent years, DNNs integrated with NURBS-based isogeometric parameterization have been used to solve forward problems in biomechanics \cite{balu2019deep,zhang2021simulating}. More recently, Liao \etal \cite{liao2022deep} developed DNN-based surrogate models for inverse material design with NURBS geometry parameterization to accelerate the traditional numerical optimization schemes. NURBS-based geometry parameterization enables dimensionality reduction and represents any shape with a fixed number of geometry parameters, hence providing an explicit approach to \textit{geometric regularization} of shapes. However, a significant concern for the DNN-based surrogate model with NURBS in \cite{liao2022deep} is its continuing dependence on traditional numerical solvers.

The existing literature highlights the need for an end-to-end deep learning based inverse design approach to uniquely solve a high-dimensional, nonlinear, and ill-posed inverse design problem. In addition, the existing DNN-based inverse design approaches also lack \textit{shape generalization} ability. For a trained neural network, shape generalization is the ability to predict any arbitrary shape within the design space. In a data-driven approach, the design space is represented by the distribution of the DNN training samples and therefore, the shape generalization ability of the data-driven DNN is essentially its \textit{in-distribution generalization} (alternatively called \textit{interpolation}) capability.  

In the present study, we introduce a  semi-supervised deep learning model for inverse design to address the limitations of the existing data-driven inverse methods. The effectiveness of this deep learning approach will be illustrated in the context of acoustic scattering inverse problems. More specifically, we develop a physics-embedded DNN with geometric regularization to design a sound hard scatterer shape capable of molding an incoming plane wave into a predefined target acoustic wave field. In the following, we will refer to this deep neural network for \textbf{G}eometric \textbf{R}egularization-based \textbf{I}nverse \textbf{D}esign of \textbf{S}hapes as \textbf{GRIDS-Net}.  

Note that, while the use of reduced order modeling and physics-based regularization approaches have been prevalent in the inverse design literature, to the best of the authors' knowledge, the introduction of geometric regularization to solve the inverse design problems via an end-to-end deep learning approach is unique to this study.
The major contributions of this study are threefold:
\begin{enumerate}
    \item \textit{Shape generalization}: The existing numerical and deep learning based inverse design methodologies approach the inverse problem by enforcing geometry constraints on the design space. Unlike these methodologies, the proposed approach does not enforce any geometry constraints and does not limit the inverse design space. As a result, GRIDS-Net can effectively predict any arbitrary scatterer shape, hence exploring a vast inverse design space.
    
    \item \textit{Geometric and physics-based regularization}: The GRIDS-Net uses NURBS geometry parameterization and problem-specific physical knowledge to regularize the ill-posed scatterer shape design problem. In order to implement the physics, an algorithm based on the boundary element method (BEM), used to evaluate the pressure field on the scatterer boundary, is developed. This algorithm is integrated within the DNN to introduce a physical observational bias in the form of additional pressure data. The BEM algorithm built into the GRIDS-Net is a key factor in regularizing the ill-posed problem. Traditional solvers embed geometric constraints to regularize the inverse design, thereby limiting the design space.  In contrast, GRIDS-Net integrates knowledge of the geometry within the DNN  and regularizes the ill-posed nature of the design problem while exploring a vast design space.
    
    \item \textit{Adaptability}: The development of BEM-integrated DNN enables rapid re-training of the GRIDS-Net to learn the inverse mapping for any boundary condition. 
    The integrated BEM algorithm considerably lowers the data-generation cost for training the GRIDS-Net. More specifically, the BEM algorithm evaluates the additional data in the form of scatterer boundary pressure based on the knowledge of the input (i.e. the incident wave). Although the numerical evaluation of the additional pressure data on the scatterer boundary might not seem a complicated task, in inverse problems of practical interest only the target pressure field is typically known, while the pressure field on the scatterer boundary remains inaccessible. 
    It is also important to highlight that the existing physics-embedded data-driven approaches incur the cost of evaluating the additional training data to enforce physical observational bias. Whereas, the GRIDS-Net can be easily retrained using target field data collected at arbitrary initial pressure conditions, while the condition-specific boundary data are evaluated by the built-in BEM algorithm. Currently, the GRIDS-Net can adapt to learn the mapping for planar incident acoustic waves of arbitrary amplitude and direction.
\end{enumerate}

The rest of the paper is organized as follows. Section 2 introduces the fundamental concepts and properties of the GRIDS-Net framework. Section 3 elaborates on the implementation of the GRIDS-Net for inverse scatterer design problems; specifically, this section details the forward problem, the numerical data extraction procedure, and the network training procedure. Finally, section 4 reports the results of the trained GRIDS-Net for two specific applications 1) remote sensing and 2) inverse material design.  

\section{Problem description}
\label{ssec: problem_setup}

This section first introduces the general problem setup of an inverse design of a scatterer in an acoustic domain and discusses the major challenges. Then, it provides a brief overview of the well-established concepts of convolutional autoencoder (CAE), non-uniform rational B-spline (NURBS), and boundary element method (BEM) in order to facilitate the understanding of the role they play to achieve reduced order model, as well as geometric and physics-based regularization. Finally, this section will conclude by defining the problem statement and present the GRIDS-Net model to solve the inverse scatterer design.

\subsection{Inverse shape design of a 2D acoustic scatterer}
\label{ssec: inverse_design_setup}

\begin{figure}[h!]
	\centering	\includegraphics[width=1.0\linewidth]{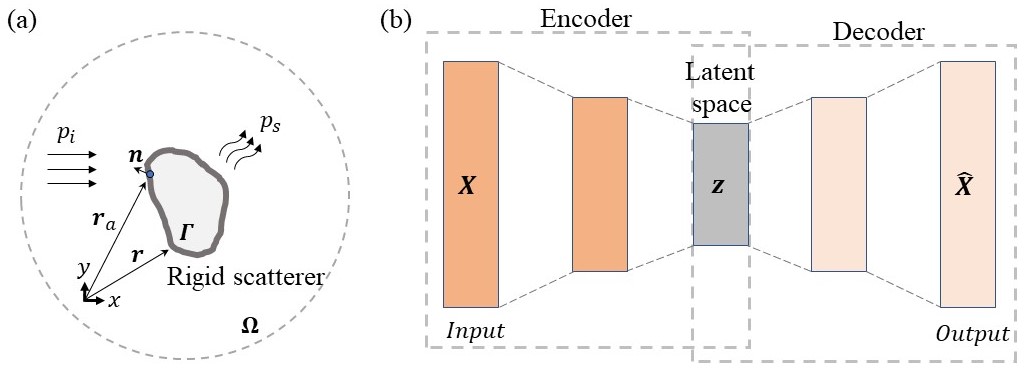}
	\caption{\label{fig: GRIDS-Net_architetcure} Schematic illustrating (a) the acoustic domain $\Omega$ and the rigid scatterer $\Gamma$. Here, $p_i$ is the incident plane pressure wave, $p_s$ is the scattered pressure wave, $\mathbf{r}$ is the position of the source of scattered wave on the acoustic scatterer, $\mathbf{r}_a$ is the position on the acoustic scatterer (blue dot), and $\textbf{n}$ is the surface normal at position $\mathbf{r}_a$. (b) The general architecture of the convolutional autoencoder (CAE). The latent space representation (\textbf{Z}) of the input data is a key feature of CAE.}
    \label{fig: AcousticDomain_and_CAE}
\end{figure}

Consider a classical steady state acoustic problem governed by the Helmholtz equation and defined on a 2D domain $\Omega \subset \mathbb{R}^2$  as shown in Fig.~\ref{fig: AcousticDomain_and_CAE}(a)
\begin{equation}
    \label{eqn: Helmholtz_equation}
    \nabla^2 p(\textbf{x}) + k^2 p(\textbf{x}) = 0, ~~~~~~~ \textbf{x}=(x,y) \in \Omega
\end{equation}
The domain includes a rigid scatterer delimited by a rigid boundary $\Gamma$ and a Sommerfeld radiation condition at infinity (open boundary)
\begin{equation}
    \label{eqn: rigid_BC}
     \frac{\partial p(\textbf{x})}{\partial \textbf{n}}=0, ~~~~~~~ \textbf{x} \in \Gamma
\end{equation}
\begin{equation}
    \label{eqn: Sommerfield_equation}
    \lim_{|\textbf{x}| \rightarrow \infty} |\textbf{x}| \Big(\frac{\partial p(\textbf{x})}{\partial |\textbf{x}|} + i k p(\textbf{x}) \Big) = 0
\end{equation}
where the total pressure $p=p_i+p_s$ is the sum of incident pressure $p_i$ and scattered pressure $p_s$, the wavenumber $k = \frac{2\pi f}{c_s}$ is a function of frequency $f$ and speed of sound $c_s$, and $\textbf{n}$ is the surface normal. $p= Re(p) + iIm(p)$ is the solution of Eq.~(\ref{eqn: Helmholtz_equation}) and is determined by the pressure scattered ($p_s$) by the rigid scatterer boundary $\Gamma$. More specifically, given a prescribed incident wave, the solution of  Eq.~(\ref{eqn: Helmholtz_equation}) is controlled by the shape of the rigid scatterer, which is the quantity of interest for the inverse design problem.

In an inverse rigid scatterer design problem, traditional numerical schemes \cite{molesky2018inverse, pestourie2018inverse} or deep learning approaches \cite{lu2021physics} can be leveraged to solve for $\Gamma$ by minimizing a PDE-constrained function that depends on both $p$ and $\Gamma$. Nevertheless, in the absence of such a direct relationship between $p$ and $\Gamma$ (or any other parameterized form of $\Gamma$ in Eq.~(\ref{eqn: Helmholtz_equation})-(\ref{eqn: Sommerfield_equation})) a new approach is needed solve the design problem.

The major challenges associated with this design problem are the high dimensionality and the strong nonlinearity of the design space, and the ill-posed nature of the inverse design. While the ability to learn highly complex nonlinear correlations in the absence of robust analytical formulations makes the DNN-based models an ideal approach to capture the nonlinearity in the design space, the use of geometry parametrization methods ensures a parsimonious representation of the high-dimensional scatterer shape. Although the use of DNNs with geometry parameterization addresses the difficulties introduced by nonlinearity and high dimensionality of the design space, the ill-posed nature of the problem remains a significant challenge. The overall design problem can be broken up into the following steps: step 1-- defining the target pressure field; step 2-- defining the geometry of the overall problem; and step 3-- defining the underlying mathematical model to capture the governing physics. In each of these steps, ill-posedness can surface due to the lack of a well-defined mathematical formulation of the governing physical problem and, more importantly, the absence of regularization approaches. Based on Eq.~(\ref{eqn: Helmholtz_equation})-(\ref{eqn: Sommerfield_equation}), the existence of a well-defined mathematical model for inverse acoustic scattering indicates that the first scenario is not a concern in this study. To address the latter concern, for step 1, we use a data-driven convolutional autoencoder to develop a parsimonious latent space representation of the target fields; step 2, we introduce geometric regularization via NURBS; and step 3, we present physics-based regularization using BEM. The geometric regularization, when supported by the physics- and the data-driven regularizations, results in a robust inverse design scheme, where each component provides a natural strategy to address potential sources of ill-posedness in the inverse design problem.           

\subsection{Reduced order modeling based on convolutional autoencoder (CAE)}
\label{ssec: CAE}

As reduced-order deep learning models are essential when dealing with high-dimensional target field-based image data, we propose a CAE as the baseline learning approach for this design problem.
Typically, CAEs are unsupervised learning models composed of convolutional layers capable of creating compressed image representations. As a result, CAEs are widely used as efficient neural network-based dimensionality reduction models capable of forming abstract representations of input images by filtering out redundant information and thereby capturing essential features. The general CAE model, as shown in Fig.~\ref{fig: AcousticDomain_and_CAE}(b), consists of three major components: 1) the encoder module, 2) the latent space representation, and 3) the decoder module. The encoder module comprises deep convolutional layers that compress the input image data into an encoded lower-dimensional representation without significant loss of information. These compressed data correspond to the latent space representation of the image input. The reduced-order latent space form of the target field data achieves a parsimonious representation of the design problem that has significant implications for the data-driven learning process. Finally, the decoder module with deep deconvolutional layers decompresses the latent space representation to reconstruct the image input, i.e. the target field.


\subsection{Geometric regularization via non-uniform rational B-spline (NURBS)}
\label{ssec: NURBS}

As most DNN architectures, including CAEs, have difficulties training with high-dimensional data, we explore the possibility for geometry parameterization of high-dimensional rigid scatterer shapes; the intent is to create its corresponding low-dimensional representation without significant loss in the accuracy of the true shape of the rigid scatterer. Moreover, the introduction of a geometry parametrization method can help regularize the ill-posed design problem. 

The concept of non-uniform rational B-spline (NURBS) has been extensively used to achieve parametric representation for various 2D geometries ranging from simple curves (e.g. circles and ellipses) to complex free-form shapes due to their ability to provide flexible and precise geometry parameterization of any arbitrary shape. More specifically, NURBS parameterization allows the mathematical description of a variety of shapes using a fixed set of control points ($\textbf{C}$) and corresponding weights ($w$). In the remainder of this section, we will summarize the key mathematical relations that allow mapping geometry coordinates to the NURBS parameters and then highlight the advantages of using NURBS for deep learning based inverse problems to design rigid scatterers. 

In order to define a NURBS curve of degree $\tilde{k}$, let us consider $n+1$ control points $[\textbf{C}_0, \textbf{C}_1,...\textbf{C}_n]$ associated with $n+1$ weights $[w_0, w_1,...w_n]$, and a knot vector $t=[t_0, t_1,...t_{m}]$ having $m+1$ knots (elements of the knot vector $t$), where $m=n+\tilde{k}+1$. Here, the knot vector is a non-decreasing sequence of geometry coordinates ($t_i$) that divides the B-splines into non-uniform piecewise functions. The NURBS curve $\tilde{\textbf{C}}$, parametrized using $u$, is defined as follows
\begin{equation}
    \label{eqn: NURBS_CP}
    \tilde{\textbf{C}}(u) = \frac{\sum_{i=0}^n w_i N_{i,\tilde{k}}(u)\textbf{C}_i}{\sum_{j=0}^n w_j N_{j,\tilde{k}}(u)}
\end{equation}
where for a $2D$ shape $\textbf{C}_i=[C^x_i, C^y_i]$ with $C^x_i$ and $C^y_i$ indicating the $x$ and $y$ coordinates of the $i^{th}$ control point, and $N_{i,\tilde{k}}(u)$ indicating the corresponding B-spline basis function.
Furthermore, based on $t$, $N_{i,\tilde{k}}(u)$ is evaluated recursively using Cox-de Boor recursion formula as follows
\begin{equation}
    \label{eqn: NURBS_basis}
    N_{i,\tilde{k}}(u) = \frac{(u-t_i)N_{i,\tilde{k}-1}(u)}{t_{i+\tilde{k}}-t_i} + \frac{(t_{i+\tilde{k}+1}-u)N_{i+1,\tilde{k}-1}(u)}{t_{i+\tilde{k}+1}-t_{i+1}}
\end{equation}
and 
\begin{equation}
    \label{eqn: NURBS_basis0}
    N_{i,0}(u)=
    \begin{cases}
        1 & t_i \leq u \leq t_{i+1}\\
        0 & \text{otherwise}
    \end{cases}
\end{equation}
For a given $t$ and $\tilde{k}$, Eq.~(\ref{eqn: NURBS_CP}) maps the NURBS parameters $\textbf{C}$ and $w$ to the NURBS curve $\tilde{\textbf{C}}$. 
There are multiple advantages to using NURBS-based geometry parameterization (Eq.~(\ref{eqn: NURBS_CP})-(\ref{eqn: NURBS_basis0})) to solve inverse design problems. First, NURBS can describe complex 2D scatterer shapes in a reduced parametric space. Second, the ability to describe any 2D shape with a compact set of parameters will significantly reduce the computational cost associated with the inverse problem. Finally, NURBS allows us to develop a generalized DNN architecture with a fixed number of output parameters capable of defining arbitrary complex scatterer geometry. 

Based on the understanding from section-\ref{ssec: CAE} and section-\ref{ssec: NURBS}, we anticipate that the CAE-based deep learning model with NURBS geometry parameterization can ease the challenges associated with high nonlinearity and high dimensionality of the inverse design problem. More importantly, the introduction of NURBS into CAE enforces geometric and data-driven regularization of the learning process. 

\subsection{Physics-based regularization utilizing boundary element method (BEM)}
\label{ssec: BEM}

DNNs with physics-based regularization methods are extensively used for both forward and inverse engineering problems \cite{nabian2020physics, karniadakis2021physics}. The ill-posedness of inverse design problems can be regularized by explicitly embedding all the \enquote{learnable} prior physical knowledge in the DNN. More specifically, these priors can act as regularizers that enhance the problem formulation and, subsequently, guide the learning process to converge to a stable, unique, and physically plausible inverse design solution. 
In this study, we choose to perform physics-based regularization through appropriate physical data and physics-driven loss functions. 

The availability of training data reflecting the acoustic response of rigid scatterers can be used as a mechanism to embed the fundamental principles of the complex scattering phenomenon into the DNN model. In this study, we leverage a BEM-based numerical solver to evaluate $p_s$ on the rigid scatterer boundary and use this information to supplement the target acoustic pressure field for DNN training. Although the finite element method (FEM) is an established numerical technique for acoustic scattering, the boundary element method (BEM) offers flexibility in numerical evaluation when the complex rigid scatterer is immersed in an infinite domain. More specifically, FEM is most practical when applied to the evaluation of fields within a finite size domain delimited by a closed boundary. 
In the following, we briefly discuss the essential elements of the BEM formulation and implementation.

The BEM is a numerical computational method commonly used to solve acoustic integral boundary value problems. The boundary integral equation corresponding to the rigid body acoustic scattering problem is a solution to the Helmholtz equation (Eq.~(\ref{eqn: Helmholtz_equation})). The interested reader can refer to \cite{fahy2007sound} for a full account on the BEM approach. The application of Green's second theorem to Eq.~(\ref{eqn: Helmholtz_equation}) yields the following integral equation for all points on the scatterer boundary $\Gamma$
\begin{equation}
    \label{eqn: BEM_eqn}
    p(\mathbf{r}) = \frac{1}{c(\mathbf{r})} \int_{\Gamma} \Big( p(\mathbf{r}_a)\frac{\partial G(\mathbf{r}, \mathbf{r}_a)}{\partial \textbf{n}} - G(\mathbf{r}, \mathbf{r}_a)\frac{\partial p(\mathbf{r}_a)}{\partial \textbf{n}}\Big)d\Gamma
\end{equation}
This equation is referred to as the \textit{direct boundary integral formulation}. Note that, in this study, Eq.~(\ref{eqn: BEM_eqn}) is formulated only to evaluate $p$ on $\Gamma$ with $c(\mathbf{r})=-\frac{1}{2}$. In Eq.~(\ref{eqn: BEM_eqn}), as shown in Fig.~\ref{fig: AcousticDomain_and_CAE}(a), $\mathbf{r}$ corresponds to the position of the source of scattered wave on the acoustic scatterer, $\mathbf{r}_a$ refers to the position on the acoustic scatterer, $p$ is the total acoustic pressure, $\textbf{n}$ is the surface normal, and $G(\textbf{r}, \textbf{r}_a)=\frac{i}{4}H_0^{(1)}(k|\textbf{r}-\textbf{r}_a|)$ is the \textit{free space} Green's function in 2D, where $H_0^{(1)}(.)$ is the zeroth order Hankel function of the first kind and $k$ is the wavenumber. Based on 
Eq.~(\ref{eqn: rigid_BC}) for a rigid scatterer with sound hard boundary condition, $\frac{\partial p(\mathbf{r}_a)}{\partial \textbf{n}}=0$. 

The solution provided by Eq.~(\ref{eqn: BEM_eqn}) applies to any type of acoustic scattering problem irrespective of the geometric complexity of the scatterer. However, the integration process becomes increasingly more elaborate for complex-shaped scatterers, and it typically requires a numerical approach, such as BEM. In its complete sense, BEM is a discretization-based numerical technique and follows a two-step procedure: 1) the \textit{boundary variables} ($p_s$ on the scatterer) are first evaluated at a discrete set of points on $\Gamma$, and 2) the \textit{field variables} ($p_s$ in target regions surrounding the scatterer) are evaluated inside the $\Omega$ domain. However, in this study, Eq.~(\ref{eqn: BEM_eqn}) evaluates the boundary variables on $\Gamma$ via BEM, while FEM-based numerical simulations evaluate the field variables inside the $\Omega$ domain. Here, the pressures on the scatterer correspond to the boundary variables and the target acoustic pressure fields are field variables.

In an inverse scattering problem, the target pressure field is known, while the pressure on the scatterer boundary is unknown due to the absence of data on the final geometry of the scatterer. Hence, the ability to regularize the ill-posed inverse design problem by embedding the DNN with information on the pressure on the scatterer (boundary variable) will significantly improve shape prediction accuracy. The following section will present a DNN architecture embedded with information regarding the boundary variables using BEM, thereby enforcing the governing physics of acoustic scattering via physical data.     

\subsection{Development of GRIDS-Net}
\label{ssec: development_of_GRIDS_Net}

Sections \ref{ssec: inverse_design_setup}-\ref{ssec: BEM} have introduced a variety of concepts with the potential to address important challenges in an inverse design problem. In order to make efficient use of these techniques, their integration into a deep learning framework is necessary.

\begin{figure}[h!]
	\centering
	\includegraphics[width=1.0\linewidth]{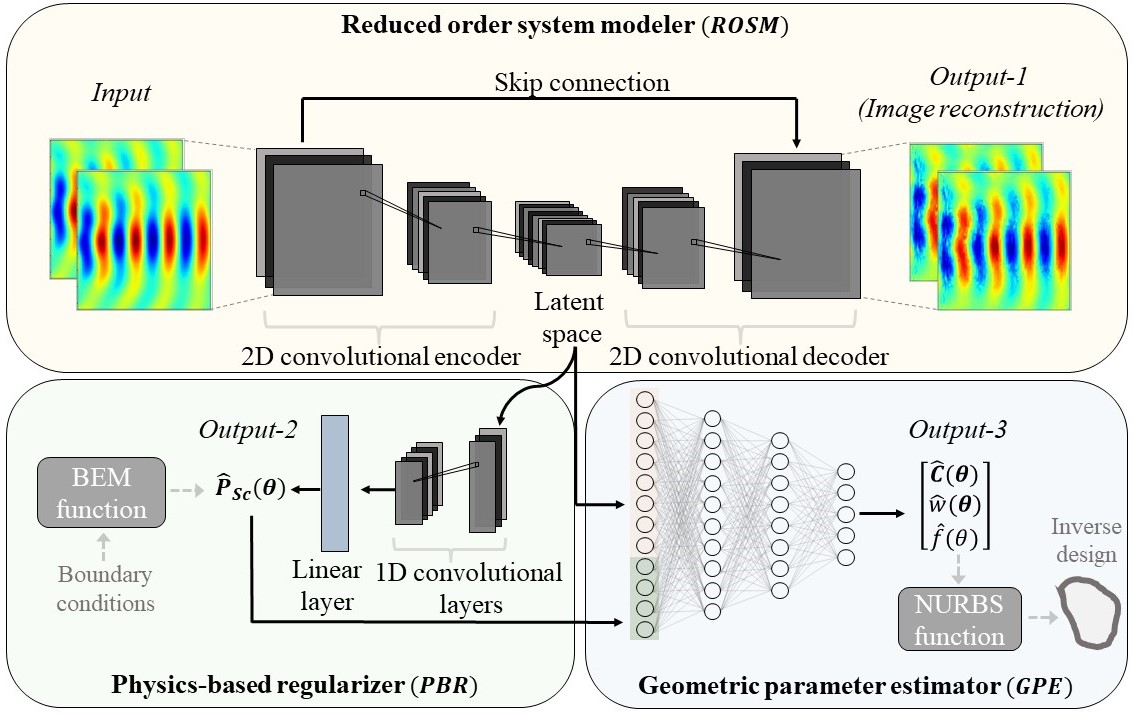}
	\caption{\label{fig: GRIDS-Net_architetcure} Schematic illustration of the architecture of the GRIDS-Net. The ROSM, PBR, and GPE modules are highlighted. Input is the target pressure field, Output-1 is the reconstructed target pressure field, Output-2 is the pressure on the scatterer ($\hat{P}_{Sc}$), and Output-3 is our primary quantity of interest for the inverse design, namely, the NURBS parameters ($\mathbf{\hat{C}}$ and $\hat{w}$) and the wave frequency ($\hat{f}$).}
    \label{fig: PE_CAE}
\end{figure}

We present a semi-supervised learning-based DNN called GRIDS-Net by integrating the key concepts of CAE, NURBS, and BEM for acoustic scattering. GRIDS-Net comprises three key modules: 1) a reduced order system modeler, 2) a physics-based regularizer, and 3) a geometric parameter estimator. Fig.~\ref{fig: PE_CAE} 
is a schematic representation of the GRIDS-Net with its modules. Hereafter, we identify and highlight the functions of the three modules as follows

\begin{enumerate}
    \item \textit{Reduced order system modeler (ROSM) module}: The ROSM module is a customized CAE that produces lower dimensional representations of high dimensional inputs. More specifically, this unsupervised learning-based CAE can produce a latent space representation of the high dimensional target pressure fields (field variables) input without significant loss in the key features of the target fields. For this study, the accuracy of the latent space representation is of utmost significance while the accuracy of the target pressure fields predicted by the ROSM module is less significant. In other words, the number of trainable network parameters in the ROSM module is controlled by the accuracy of the reduced order latent space representation of the target fields rather than by the accuracy of the target field reconstruction. This is primarily due to the significant role the latent space plays in the development of a parsimonious correlation with the parameterized inverse design.
    
    \item \textit{Physics-based regularizer (PBR) module}: The PBR module enforces the physical knowledge of the design problem into the GRIDS-Net. The PBR module takes the latent space representation from the ROSM module as input to predict the pressure on the scatterer, $\hat{P}_{Sc}$. A BEM-based numerical solver is integrated into the PBR module to provide the true data. Specifically, the BEM solver allows solving Eq.~(\ref{eqn: BEM_eqn}) to evaluate the true pressure on the scatterer (boundary variable), $P_{Sc}$. This robust solver calculates $P_{Sc}$ using the properties of the incident acoustic wave and the discretized geometry of the scatterer. The BEM solver is indicated by the box \enquote{BEM function} in Fig.~\ref{fig: PE_CAE}. It is important to highlight that the BEM solver is not part of the training process and has no trainable parameters; the solver is only responsible for providing the true data to the PBR module. The development of an integrated BEM solver is motivated by the practical limitation of directly measuring pressure on complex scatterer shapes. 
    Further, the flexibility of this BEM-integrated PBR module allows the users to easily retrain the network for arbitrary boundary conditions. In the current scenario, the BEM solver is developed for an incident acoustic wave of arbitrary amplitude and direction.

    \item \textit{Geometric parameter estimator (GPE) module}: The GPE module is responsible for shape prediction via geometric regularization. It has two sub-components: 1) a fully connected neural network (FCNN), and 2) a NURBS curve calculator. The first and the most important sub-component, the FCNN, maps the target pressure fields to the parameterized inverse scatterer geometry. More specifically, the FCNN acts as a supervised learning model that takes the concatenation of the latent space representation from the ROSM module and $\hat{P}_{Sc}$ from the PBR module as input to predict NURBS parameters, $\hat{\textbf{C}}$ and $\hat{w}$, as its output. Additionally, the GPE module also predicts the wave frequency, $\hat{f}$, of the target field. The second sub-component, the NURBS curve calculator, evaluates the scatterer geometry coordinates using $\hat{\textbf{C}}$ and $\hat{w}$ by coupling the NURBS formulation (Eq.~(\ref{eqn: NURBS_CP})-(\ref{eqn: NURBS_basis0})) to the FCNN predictions. The NURBS curve calculator is indicated by the box \enquote{NURBS function} in Fig.~\ref{fig: PE_CAE}. Similarly to the BEM solver in the PBR module, the NURBS curve calculator has no trainable parameters and is used to evaluate the scatterer geometry from $\hat{\textbf{C}}$ and $\hat{w}$.

\end{enumerate}

The above description suggests that, at least at a conceptual level, the GRIDS-Net could solve the inverse scatterer design problem by addressing some fundamental challenges (high-dimensionality, nonlinearity, and ill-posedness) without constraining the inverse design space. Additionally, for specific boundary conditions, GRIDS-Net is capable of shape generalization without retraining the neural network.

\section{Methodology} 
\label{ssec: Methodology}

In this section, we focus on the setup and implementation of the GRIDS-Net. In the context of developing a deep learning based inverse design model, the first task is to set up a forward acoustic scattering model to generate the scattering response based on a set of scatterer shapes and boundary conditions. Subsequently, the objective of the inverse design task is to predict the scatterer shapes based on the predefined target scattering response. In a broad sense, we define the exact forward model used for numerical data generation and then study the GRIDS-Net architecture and its training process for the inverse design.  

\subsection{Forward problem}
\label{ssec: forward_problem}

This section introduces the forward acoustic scattering simulation as a two-stage process. The first stage involves determining the NURBS-based scatterer geometry, while the second stage defines the numerical simulation setup for the scatterer geometry determined in the first step. The remainder of this section will introduce the dataset generation process by utilizing the forward simulation model. The schematic in Fig.~\ref{fig: geometry_sample_INV} represents a prototypical geometry and associated boundary conditions used in the numerical simulations.

\subsubsection{NURBS-based scatterer shape}

The rigid scatterer shapes in the forward numerical simulations are defined via NURBS. While section \ref{ssec: NURBS} introduced the concept of NURBS and its importance in this study, this particular section is dedicated to the details regarding its implementation. The work by Wu \etal \cite{wu2022physics} was used as a basis to determine the range of NURBS parameters. All the NURBS-based scatterer shapes are represented with a knot vector t=[0,0,0,1/4,1/4,1/2,1/2,3/4,3/4,1,1,1] and quadratic B-spline basis functions. In this study, we select nine control points $[\textbf{C}_0, \textbf{C}_1,..\textbf{C}_8]$, and corresponding weights $[w_0, w_1,...w_8]$ to represents all shapes; this range of NURBS parameters seems to provide enough flexibility to capture a broad range of shapes \cite{piegl1996nurbs, wu2022physics}. As the scatterer shapes of interest are represented by closed curves, the first and last NURBS control points $\textbf{C}_0$ and $\textbf{C}_8$, respectively, overlap. Therefore, the total number of unknowns necessary to represent any arbitrary scatterer shape with NURBS is a combination of eight control points in $x$-direction $[C^x_0, C^x_1,...C^x_7]$ and eight control points in $y$-direction $[C^y_0, C^y_1,...C^y_7]$ and the corresponding weights $[w_0, w_1,...w_7]$, i.e. a total of 24 NURBS parameters for a 2D scatterer shape. Due to the possibility of multiple design strategies, in this study, we impose a numerical range for $\textbf{C}$ and $w$. Firstly, we define an upper limit on the characteristic length $l_0^c=2l_0$ of the scatterer, where $l_0$ is the distance between the center of the scatterer and the farthest $\textbf{C}$ on a given scatterer (Fig.~\ref{fig: geometry_sample_INV}(a)). Further, $l_0^{max}$ is the maximum value of $l_0$ among all possible scatterers, i.e. $l_0^{max}=max(l_0)$. The upper limit $l_0^{max}$ is defined such that no scatterer shape can occupy more than 10\% of the total area of the domain as shown in Fig.~\ref{fig: geometry_sample_INV}(a). Secondly, we introduce a lower limit $l_0^{min}$ as the distance between the center and the nearest $\textbf{C}$ among all the scatterer designs such that below $l_0^{min}$ the 2D NURBS shapes no longer remain a simple closed curve. In the current design scenario, $(2l_0^{min}, 2l_0^{max})$=(0.1, 0.3) $m$ and lastly, based on the observation from \cite{wu2022physics}, $w$ is bounded within the range $[0,1]$ as the variation in the shape is minimal for $w$ values greater than $1$. Table~\ref{table:1} reports the numerical values of the key parameters used to define the NURBS-based scatterer geometry. It is important to note that although the NURBS parameters are bounded, the number of possible inverse designs is infinite as $C^x_i, C^x_i, w_i \in \mathbb{R}^b \subset \mathbb{R}$, where $\mathbb{R}^b$ is the bounded real space of the NURBS parameters and $i=0,2,...7$.

\begin{table}[h!]
\setlength\tabcolsep{0pt}
	\begin{center}
		\begin{tabular}{ |c|c|c| } 
			\hline
			\hline
			\multirow{1}*{\textbf{Parameter}} &  \textbf{Symbol} & \textbf{Numerical value}\\
			\hline
			\hline
			Degree of NURBS curve & $\tilde{k}$ & 2 \\
			\hline   
			Number of control points & $n+1$ & 9 \\
            \hline   
			Number of knots & $m+1$ & 12 \\
			\hline        
			Knot vector ($m$) & $t$ & [0, 0, 0, $\frac{1}{4}$, $\frac{1}{4}$, $\frac{1}{2}$, $\frac{1}{2}$, $\frac{3}{4}$, $\frac{3}{4}$, 1, 1, 1] \\
			\hline
			Control points ($m$)& $\textbf{C}=[C^x, C^y]$ & $|C^x| \in [0.05,0.15],~  |C^y| \in [0.05,0.15]$\\
			\hline
			Weights & $w$ & $w \in [0,1]$\\
			\hline
			Area of acoustic domain ($m^2$) & $A=a \times b$ & $1 \times 1$\\
			\hline
   		Distance between scatterer boundary 
            & $\Delta d$ &  0.0225\\
            and measurement domain ($m$)& & \\
			\hline
		\end{tabular}
		\caption{Summary of key geometry parameters essential for the development of the NURBS-based rigid scatterer design immersed in an acoustic domain. The Schematics in Fig.~\ref{fig: geometry_sample_INV} highlight some of these geometry parameters.}
		\label{table:1}
	\end{center}
\end{table}

\subsubsection{Simulation setup}

\begin{figure}[h!]
	\centering
	\includegraphics[width=0.95\linewidth]{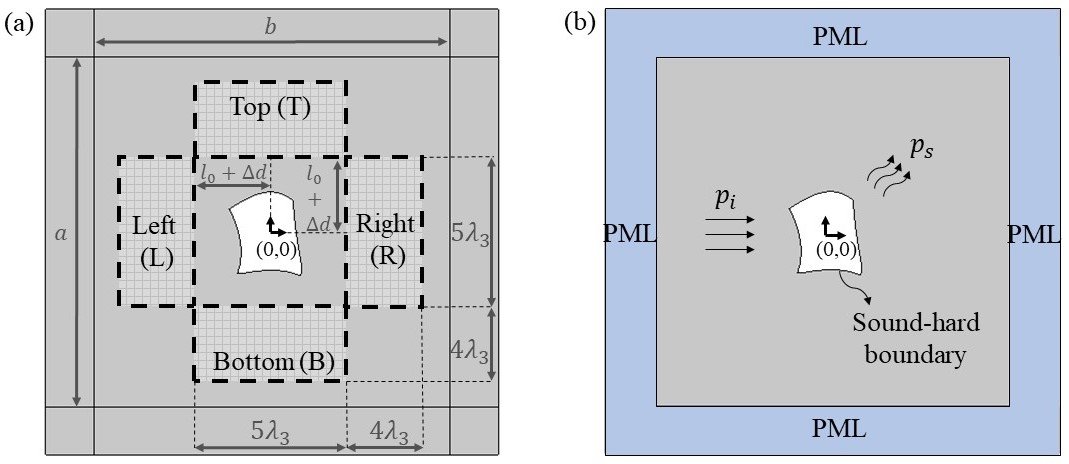}
	\caption{ (a) Schematic illustrating the scatterer surrounded by the measurement domains (shown by the patterned areas) L, T, R, and B. The measurement domain is fixed and is specified for wavelength ($\lambda_3$) corresponding to $f= 5~kHz$. Note that the physical coordinates of the measurement domain remain constant at all frequencies. (b) Schematic depicting the incident acoustic wave ($p_i$), scattered acoustic wave ($p_s$), and the boundary conditions for the forward numerical simulation.
 }
	\label{fig: geometry_sample_INV}
\end{figure}

The forward simulation setup considers an acoustic plane wave $p_i$ of unit amplitude, travelling in the $x$-direction, incident on a single rigid scatterer located at the center of the domain as shown in Fig.~\ref{fig: geometry_sample_INV}(b). Due to the choice of open boundaries (Eq.~(\ref{eqn: Sommerfield_equation})), the rigid scatterer geometry controls the acoustic scattering in the infinite domain. We replicate open boundaries by enforcing artificial energy-absorbing conditions through perfectly matched layers (PMLs) on the external boundaries, as shown in Fig.~\ref{fig: geometry_sample_INV}(b). Further, a sound-hard internal boundary condition (Eq.~(\ref{eqn: rigid_BC})) is enforced on the rigid scatterer. Based on $l_0^c$, the forward simulations following a planar incident wave input are performed at three different frequencies (or wavelengths $\lambda$) corresponding to $kl_0^c<1$ ($\lambda_1$), $kl_0^c \approx 1$ ($\lambda_2$), and $kl_0^c>1$ ($\lambda_3$), where the wavenumber $k = \frac{2\pi f}{c_s}$ is a function of frequency $f$ and speed of sound in air $c_s=343.21~m/s$. More specifically, the forward model is simulated at frequencies $f_1=100~Hz$ $(\lambda_1=3.4321~m)$, $f_2=500~Hz$ $(\lambda_2=0.6864~m)$, and $f_3=5~kHz$ $(\lambda_3=0.0686~m)$. While the low frequencies (long wavelengths) scattered pressure fields capture information regarding the equivalent area of the scatterer, the high frequencies (or short wavelengths) scattered pressure fields capture information regarding the finer geometric details of the scatterer's boundary. The above explains the choice of the different frequency values.
%
In this study, all the numerical simulations based on the Helmholtz equation (Eq.~(\ref{eqn: Helmholtz_equation})) were performed in COMSOL Multiphysics$\textsuperscript{\textregistered}$.

\subsubsection{Dataset generation}
\label{ssec: data_extraction}

Based on the forward model introduced in the previous sections, we perform acoustic scattering simulations for different scatterer shapes and gather data to train the GRIDS-Net. Fig.~\ref{fig: surrogate_model}(a) shows the scattered pressure field evaluated using the forward model for a sample scatterer shape. In $\mathbb{R}^b$ as $|C^x| \in [0.05,0.15]~m$, $|C^y| \in [0.05,0.15]~m$, and $w \in [0,1]$, all the 24 NURBS parameters are uniformly distributed using Latin hypercube sampling (LHS) technique to generate 1750 sample scatterers within the ranges of $C^x$, $C^y$, and $w$. LHS is an effective sampling technique for generating uniformly distributed random samples of parameters from multidimensional data.

For each scatterer shape and frequency of excitation, the real and imaginary parts of the scattered pressure in the target windows are recorded. The current study chooses to record the pressure on four target windows, namely Left (L), Top (T), Right(R), and Bottom (B), each with a grid size $210 \times210$, surrounding the scatterer as highlighted in Fig.~\ref{fig: geometry_sample_INV}(a) and Fig.~\ref{fig: surrogate_model}(a). All the target windows have a fixed domain size (5$\lambda_3 \times$4$\lambda_3$) and grid size to capture the shortest wavelength ($\lambda_3=0.0686~m$). In total, we record 24 target pressure fields (4 target windows $\times$ 3 frequencies $\times$ 2 pressure components (that is real and imaginary)).

\subsection{Inverse design: GRIDS-Net architecture and training}
\label{ssec: Architecture_Train}

Section \ref{ssec: development_of_GRIDS_Net} broadly introduced the GRIDS-Net framework and highlighted the key modules and their functions for the design problem. This section first presents a detailed discussion of the GRIDS-Net architecture with emphasis on the specific neural layers used in these modules. Then, it will focus on the GRIDS-Net training process.

\subsubsection{Network architecture}

As introduced in Section \ref{ssec: development_of_GRIDS_Net}, the GRIDS-Net architecture consists of three key modules, namely ROSM, PBR, and GPE modules, as shown in Fig.~\ref{fig: GRIDS-Net_architetcure}. While the overarching functions of the individual modules were highlighted earlier, a detailed discussion on the network architectures of the individual modules is presented below. 

\begin{enumerate}
    \item The ROSM module is a CAE (see Fig.~\ref{fig: GRIDS-Net_architetcure}) that consists of an encoder and a decoder setup. The encoder consists of three 2D convolutional layers and two max pooling layers with batch normalization and zero padding. The number of channels in the three convolutional layers is 64, 16, and 8, respectively. Similarly, the decoder has three 2D convolutional transpose layers and two upsampling layers with batch normalization and zero padding.  The number of channels in the three convolutional transpose layers is 8, 16, and 64, respectively. Each layer in the CAE network is followed by a \textit{ReLU} activation layer, and  all the convolution operations are performed with a constant kernel of size 3. In addition, a skip connection addresses the vanishing gradient issue and forces faster convergence. The input to the ROSM module (marked $Input$ in Fig.~\ref{fig: GRIDS-Net_architetcure}) is the target acoustic pressure fields surrounding the scatterer. While the primary output of the ROSM module (marked $Output-1$ in Fig.~\ref{fig: GRIDS-Net_architetcure}) is the reconstructed input, an accurate latent space representation of the input is our quantity of interest.

    \item The PBR module consists of three 1D convolutional layers and a linear layer. Each 1D convolutional layer is followed by a $ReLU$ activation layer with batch normalization and zero padding. All the convolution operations are performed with a constant kernel of size 5. The number of channels in the three 1D convolutional layers is 8, 4, and 2, respectively, followed by a linear layer with 200 nodes. The PBR module predicts $\hat{P}_{Sc}$ (marked $Output-2$ in Fig.~\ref{fig: GRIDS-Net_architetcure}) with the latent space representation from the ROSM module as the input.

    \item The GPE module consists of a fully-connected neural network (FCNN) architecture to estimate the NURBS parameters of the scatterer shape. While the input to the FCNN is a concatenation of the vectorized latent space representation from the ROSM module followed by a dropout layer with a 0.25 dropout rate and $\hat{P}_{Sc}$ from the PBR module, the FCNN contains four hidden linear layers with 512, 512, 128, and 64 nodes, respectively. In addition, each layer in the GPE module is followed by \textit{Leaky ReLU} activation layer. The output layer of FCNN splits into three branches with linear layers to predict $\hat{\textbf{C}}$, $\hat{w}$, and $\hat{f}$ (marked $Output-3$ in Fig.~\ref{fig: GRIDS-Net_architetcure}).
\end{enumerate}

Although there are multiple independent DNNs in the modules, the GRIDS-Net is trained jointly as a hybrid network under a semi-supervised training paradigm. The GRIDS-Net is jointly trained by minimizing the following loss function
\begin{equation}
\begin{split}
     \label{eqn: loss_func}
     Loss &= Loss_{ROSM} + Loss_{PBR} + Loss_{GPE} + Loss_{Phy} \\
\end{split}     
\end{equation}
where 
\begin{equation}
\begin{split}
     \label{eqn: loss_func_comp}
     Loss_{ROSM} &= \frac{1}{N} \sum_{i=1}^N (\hat{y}_i - y_i)^2 \\ 
     Loss_{PBR} &= \frac{1}{N}\sum_{i=1}^N (\hat{P}_{{Sc}_i} - P_{{Sc}_i})^2\\
     Loss_{GPE} &= \frac{\omega_1}{N}\sum_{i=1}^N (\hat{\textbf{C}}_i - \textbf{C}_i)^2 + \frac{\omega_2}{N}\sum_{i=1}^N (\hat{w}_i - w_i)^2 + \frac{\omega_3}{N}\sum_{i=1}^N \sum_{j=1}^M -f^j_i \log (p(\hat{f}^j_i))\\ 
     Loss_{Phy} &= \frac{\omega_4}{N}\sum_{i=1}^N \Big(\Big(\frac{\partial \hat{\textbf{C}}_i}{\partial \hat{P}_{{Sc}_i}}\Big)^T \mathbb{I}_{\textbf{C}} - \textbf{D}_{P_{{Sc}_i}} \textbf{C}_i \Big)^2 + \frac{\omega_5}{N}\sum_{i=1}^N \Big(\Big(\frac{\partial \hat{w}_i}{\partial \hat{P}_{{Sc}_i}}\Big)^T \mathbb{I}_w - \textbf{D}_{P_{{Sc}_i}} w_i \Big)^2 \\  
\end{split}     
\end{equation}
where $y$ is the target pressure field, $P_{{Sc}}$ is the pressure on the acoustic scatterer, $\textbf{C}$ are the NURBS control points, $w$ are the NURBS weights, $f \in [0,1,2]$ is the $M=3$ classification label corresponding to the three acoustic wave frequencies, $p(.)$ is the probability of an event (a classification label in this case), $[\omega_1, \omega_2, \omega_3, \omega_4, \omega_5]=[2,2,0.01,100,100]$ are the network weighting factors, and $N$ is the number of training samples. While these variables represent the ground truths, the corresponding GRIDS-Net predictions are represented using the same variables with a $\hat{.}$ symbol. Note that we use regression-based \textit{mean square error (MSE)} loss for all the GRIDS-Net predictions except for $\hat{f}$. We use \textit{cross entropy (CE)} loss to predict multi-category wave frequencies as all data samples can be classified into three fixed values of $\hat{f}$. Further, the additional loss term $Loss_{Phy}$ is introduced in the total loss. $Loss_{Phy}$ acts as a soft penalty constraint that forces GRIDS-Net to converge to a design solution weakly satisfying the underlying physics. In addition to the physical information in the $P_{Sc}$ data, $Loss_{Phy}$ introduces an additional learning bias by connecting $\hat{P}_{Sc}$ with $\hat{\textbf{C}}$ and $\hat{w}$ through their respective gradients. In $Loss_{Phy}$, the ground truth $\textbf{D}_{P_{Sc}} (.)$ is the sum of the gradients with respect to $P_{Sc}$, $\mathbb{I}_{\textbf{C}}$ is an all-ones vector compatible with the transpose of the Jacobian of $\hat{\textbf{C}}$ with respect to $\hat{P}_{Sc}$, and $\mathbb{I}_w$ is an all-ones vector compatible with the transpose of the Jacobian of $\hat{w}$ with respect to $\hat{P}_{Sc}$. As the $Loss_{Phy}$ is defined as a MSE loss, it is imperative to provide the ground truth $\textbf{D}_{P_{Sc}} (.)$ to successfully evaluate the loss.            

\begin{figure}[h!]
	\centering
	\includegraphics[width=0.9\linewidth]{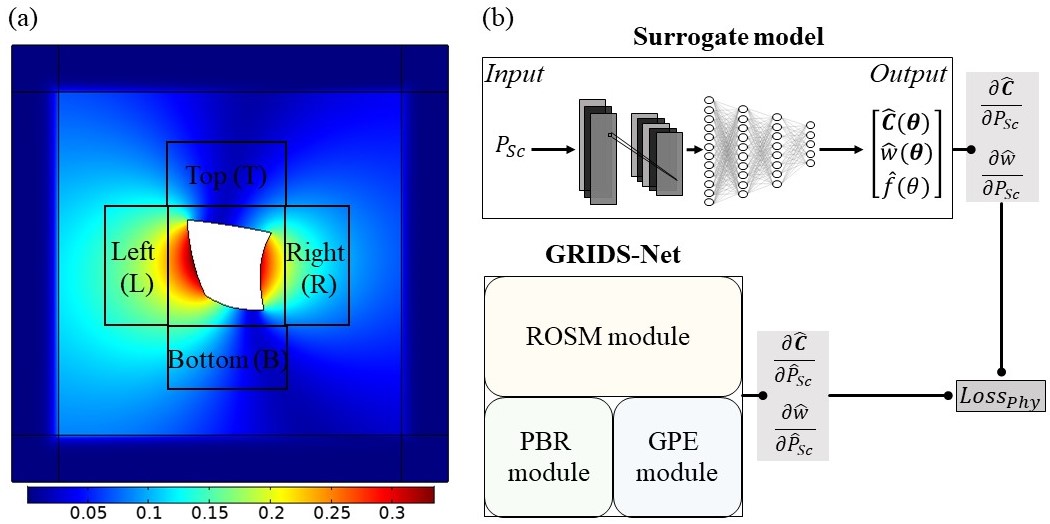}
	\caption{Schematic illustrates (a) the scattered pressure field evaluated for a sample scatterer shape using the forward model and (b) the flow of gradient information from the surrogate model and the GRIDS-Net model to evaluate $Loss_{Phy}$. The schematic in (b) also highlights the neural network architecture of the surrogate model.}
	\label{fig: surrogate_model}
\end{figure}

In this study, $\textbf{D}_{P_{Sc}} (.)$ required to calculate $Loss_{Phy}$ is evaluated via another data-driven neural network as shown in Fig.~\ref{fig: surrogate_model}(b). We develop a neural network-based surrogate model to map $P_{Sc}$ to $\hat{\textbf{C}}$ and $\hat{w}$. This surrogate model is responsible for evaluating $\textbf{D}_{P_{Sc}} \textbf{C} = (\frac{\partial \hat{ \textbf{C}}}{\partial P_{Sc}})^T \mathbb{I}_{\textbf{C}}$ and $\textbf{D}_{P_{Sc}} w = (\frac{\partial \hat{w}}{\partial P_{Sc}})^T \mathbb{I}_{w}$ values to calculate $Loss_{Phy}$ in Eq.~(\ref{eqn: loss_func_comp}). The model consists of three 1D convolutional layers of kernel size 5 with 2, 4, and 8 channels, respectively, followed by a linear layer with 64 nodes. Each layer is followed by a $Tanh$ activation layer. The output layer of the surrogate model splits into three branches to predict $\hat{\textbf{C}}$, $\hat{w}$, and $\hat{f}$. This surrogate model is trained by minimizing the following loss function 
\begin{equation}
\begin{split}
     \label{eqn: loss_func_SM}
     Loss_{SM} &= \frac{\omega_1}{N}\sum_{i=1}^N (\hat{\textbf{C}}_i - \textbf{C}_i)^2 + \frac{\omega_2}{N}\sum_{i=1}^N (\hat{w}_i - w_i)^2 + \frac{\omega_3}{N}\sum_{i=1}^N \sum_{j=1}^M -f^j_i \log (p(\hat{f}^j_i))\\ 
\end{split}     
\end{equation}
Note that $Loss_{SM}$ (Eq.~(\ref{eqn: loss_func_SM})) is comparable with $Loss_{GPE}$ in Eq.~(\ref{eqn: loss_func_comp}). Although the loss formulations are the same, the neural network architectures of the surrogate model and the GPE module in the GRIDS-Net are different. Moreover, the surrogate model develops an accurate mapping of $P_{Sc}$ to $\hat{\textbf{C}}$ and $\hat{w}$ and subsequently evaluates accurate gradients via the smaller but effective neural network architecture of the model. Once the above-mentioned surrogate model is trained to evaluate $\textbf{D}_{P_{Sc}}(.)$, the gradient information is used to calculate $Loss_{Phy}$ to train the GRIDS-Net (Fig.~\ref{fig: surrogate_model}(b)). Note that this surrogate model trains on the same dataset as the GRIDS-Net; hence, no additional training data is required. Furthermore, all the gradients required to calculate $Loss_{Phy}$ are evaluated using automatic differentiation. The optimal number and type of neural layers, activation functions, and weighting factors used in this study are chosen by trial and error.

\subsubsection{Network training}

As previously mentioned, the normalized real and imaginary parts of the scattered target pressure fields simulated at three frequencies and measured at four target windows are provided as input to the ROSM module. In order to enhance the performance of the GRIDS-Net with the available data, we perform a data augmentation process before providing the training input. The three sets of frequency data collected on each of the 1750 scatterer samples are augmented to provide a total size of 5250 (1750 $\times$ 3) samples for network training and validation. This total dataset is divided into training and validation datasets in 80:20 ratio. While the training dataset is used to fit the GRIDS-Net, the validation dataset is used to provide an unbiased model fit on the training data via hyperparameter tuning. Further, a set of 100 unseen scatterer shapes form the test dataset. The test data consists of scatterer samples that have never been known to the trained GRIDS-Net. The prediction accuracy of the GRIDS-Net based inverse design model is calculated on the test dataset.

With the augmented data, the total inputs to the ROSM module are 8 image channels of size $210\times 210 \times 8$. Based on the latent space representation input from the ROSM module, the PBR module predicts $\hat{P}_{Sc}$. The PBR module calculates pressure on 100 discrete points on the scatterer shape and stacks the real and imaginary pressure values to form a single vector of size $200 \times 1$. Further, the input to the GPE module is a reduced order latent space representation of size $21632 \times 1$ from the ROSM module concatenated with $\hat{P}_{Sc}$ of size $200 \times 1$ from PBR module. The GPE module splits in three branches to predict $\hat{\mathbf{C}}$, $\hat{w}$, and $\hat{f}$ as vectors of size $16 \times 1$, $8 \times 1$, and $3 \times 1$, respectively. Once trained, GRIDS-Net is tested for accuracy using the test dataset.

The GRIDS-Net is trained using the Adam optimizer with an adaptive learning rate of 1e-4, weight decay of 1e-4, and a batch size of 32 samples for 1000 epochs. The network is trained until the training and validation loss converge without overfitting. The proposed GRIDS-Net model is implemented in Python 3.8 using Pytorch. The NVIDIA A100 Tensor Core GPU with 40 GB memory was used to train the network.

\section{Results}
\begin{figure}[h]
	\centering
	\includegraphics[width=0.9\linewidth]{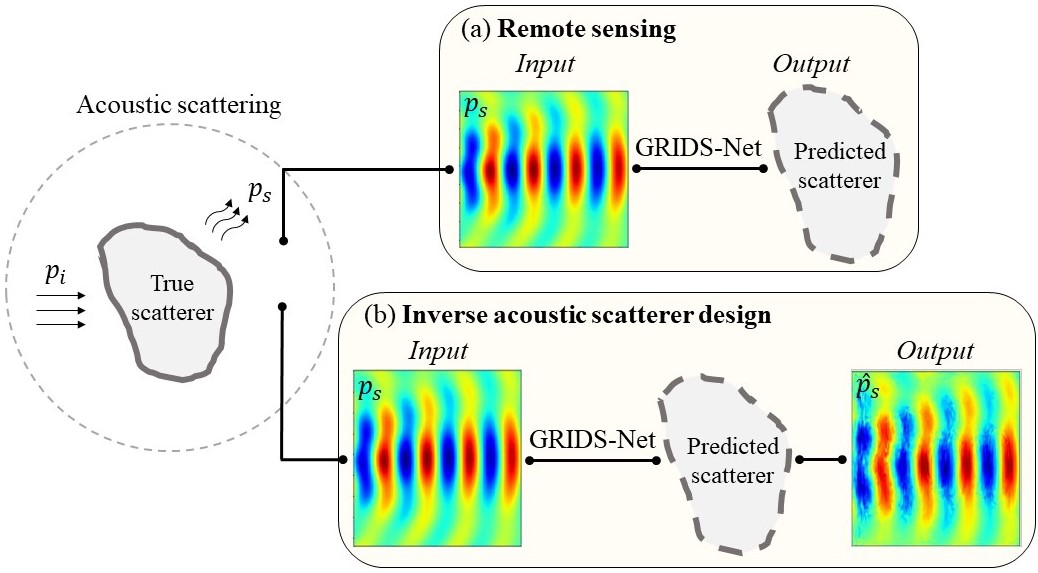}
	\caption{Schematic illustrating the use of GRIDS-NET to design acoustic scatterers for two classes of problems. (a) \textit{Remote sensing}: GRIDS-Net predicts accurate rigid acoustic scatterer shapes (marked \textit{Output}) from the knowledge of the true scattered pressure fields $p_s$ (marked \textit{Input}). This class of problems focuses on accurately capturing the scatterer shape capable of producing $p_s$. (b) \textit{Inverse acoustic scatterer design}: GRIDS-Net predicts rigid acoustic scatterer shapes to achieve accurate scattered pressure fields $\hat{p}_s$ (marked \textit{Output}) from the knowledge of true scattered pressure fields $p_s$ (marked \textit{Input}). Here, the focus is to create a scatterer design capable of reconstructing $p_s$.}
	\label{fig: applications}
\end{figure}

This section presents an extensive numerical study to explore the performance of GRIDS-Net for the inverse design of rigid acoustic scatterers. Although inverse scatterer design problems span a wide range of engineering applications, we focus on using GRIDS-Net to design rigid acoustic scatterers for two important classes of problems as shown in Fig.~\ref{fig: applications}: 1) remote sensing and 2) inverse acoustic scatterer design. 

In the context of the first class of problems, that is remote sensing, the intent is to use GRIDS-Net to predict the shape of a scatterer from the knowledge of the scattered field. Concerning the second class of problems, that is the inverse acoustic scatterer design, the intent is to use GRIDS-Net to predict the necessary shape of the scatterer to achieve a target acoustic field. While in the first study, the emphasis is on the accuracy of the reconstructed shape, in the second, the more important aspect is the accuracy of the generated target pressure field. Note that, to some extent, the second class of problems could be considered a special case of the first, as exact shape predictions correlate directly to accurate target responses. In the remainder of the paper, the scattered pressure field produced by the predicted scatterer shape is called the \textit{target acoustic field}, and the scattered pressure field produced by the true scatterer shape is called \textit{true acoustic field}.


The results for the inverse scatterer design prediction accuracy will be presented in section \ref{ssec: remote sensing}, while the results for the target acoustic field will be summarized in section \ref{ssec: material design}. Before discussing the results, it is important to introduce the different metrics used to assess the overall performance of the GRIDS-Net based inverse design approach. We introduce the following metrics 
\begin{enumerate}
    \item \textbf{Qualitative assessment:} To qualitatively assess the performance of the GRIDS-Net, we present a \textit{visual comparison} of the predicted scatterer with the true scatterer design. While the visual comparison can provide an overview of the prediction accuracy of the selected scatterer shape samples, it is necessary to introduce visual metrics that can provide an overall assessment across all the samples. For this reason, we present \textit{Quantile-Quantile} (Q-Q) plots capable of graphically comparing the probability distributions of the predicted and true parameters. If the predicted and true parameters belong to the same sample distribution, the Q-Q plot will lie on a straight line. 

    \item \textbf{Quantitative assessment:} While the qualitative assessment metrics can provide a broad overview of the performance of the design approach, it is important to introduce objective metrics that can provide a quantitative assessment of the results. In this regard, we introduce the following metrics
    \begin{enumerate}
        \item \textit{Discrete Fr\'echet distance ($\delta_f$)}: Intuitively, shape comparisons seem simple since the human brain is naturally wired to recognise common patterns in shapes. However, mathematically, comparing two shapes with unique numerical representations is more complicated. The Discrete Fr\'echet distance provides an approximate comparison of the Fr\'echet distance between two arbitrary 2D shapes \cite{eiter1994computing}. The Fr\'echet distance between the predicted and the true scatterer shape is a measure of similarity between the two shapes based on the location and order of the points along the shapes. In a discrete sense, we define the Fr\'echet distance as
        \begin{equation}
            \delta_f(P,Q):= d(i,j) = \max\Big[||p_{i(t)}-q_{j(t)}||, \min\big( d(i-1, j-1), d(i-1,j), d(i, j-1)\big)\Big]
        \end{equation}
        where $p=[p1,p2,p3,..p_n]$ and $q=[q1,q2,q3,..q_m]$ are sequence of points on the two shapes $P$ and $Q$. The Fr\'echet distance $\delta_f$ is the maximum of the minimum distances $d$ between the points $p_i$ and $q_j$ across time instances $t$.
        In this study, we use the Python implementation of the $\delta_f$ algorithm \cite{Jekel2019} to evaluate the accuracy of the inverse scatterer design predictions.  
        \item \textit{Relative error \%}: The Relative error \% metric is also introduced to measure the difference between the target acoustic field and the true acoustic field as follows
        \begin{equation}
            \text{Relative error \%} = \Big|\frac{\text{True acoustic field} -\text{Target acoustic field}}{\text{True acoustic field}}\Big| \times 100
        \end{equation}            
    \end{enumerate}
\end{enumerate}

\begin{figure}[h]
	\centering
	\includegraphics[width=1.0\linewidth]{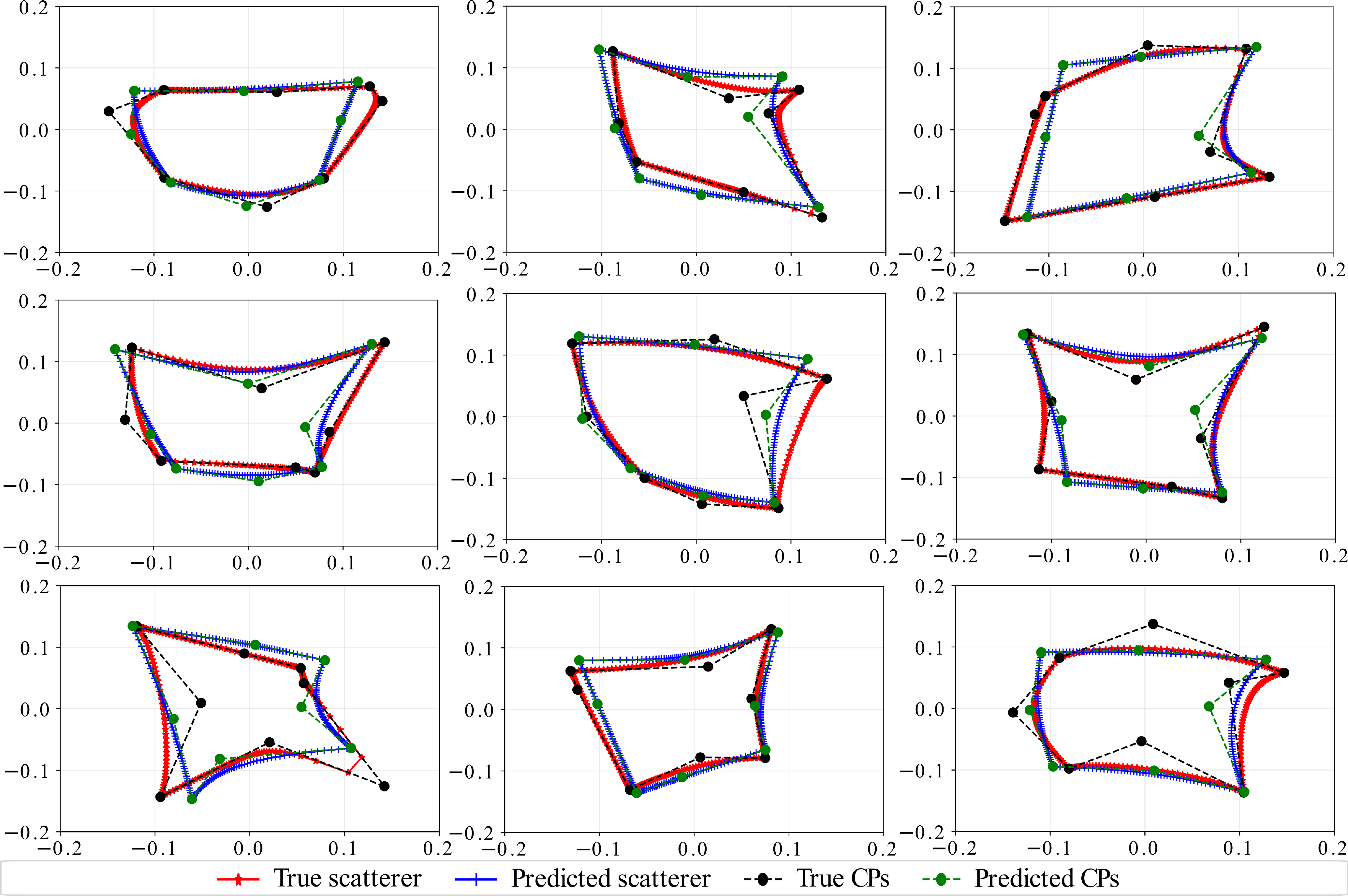}
	\caption{Visual comparison of the GRIDS-Net predicted scatterer shape (\textit{Predicted scatterer}) and the true scatterer shape (\textit{True scatterer}) for nine random samples within the test dataset. The plot also highlights the position of the eight predicted NURBS control points (\textit{Predicted CPs}) with the true NURBS control points (\textit{True CPs}) in each sample.}
	\label{fig: remote_sensing_results}
\end{figure}

\subsection{Remote sensing application: numerical results and performance assessment}
\label{ssec: remote sensing}

\begin{figure}[h]
	\centering
	\includegraphics[width=1.0\linewidth]{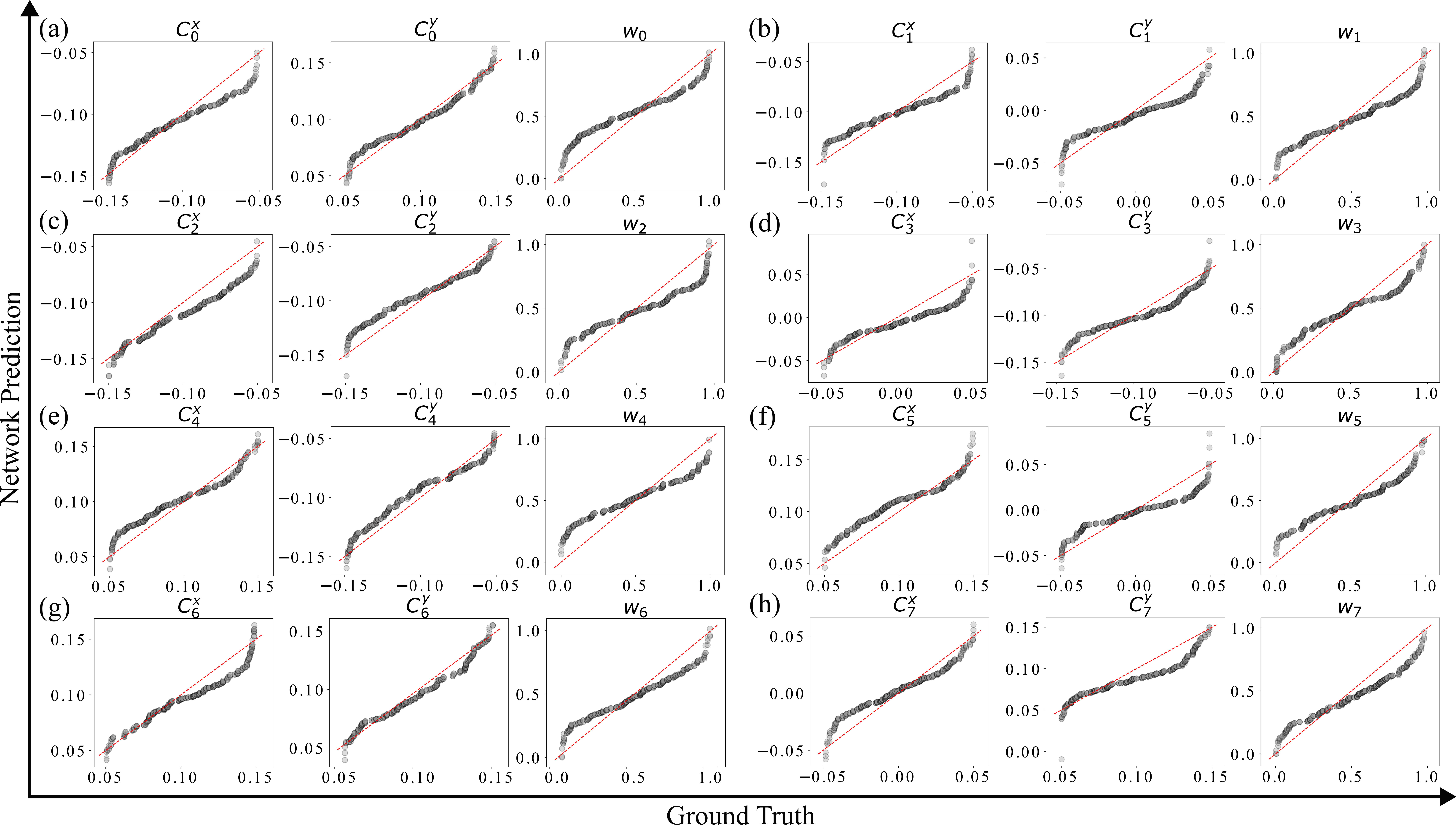}
	\caption{The Quantile-Quantile (Q-Q) plots of the GRIDS-Net predicted NURBS parameters. Each Q-Q plot compares the distribution of the predicted NURBS parameters to the distribution of true NURBS parameters for all the samples in the test dataset. Plots (a)-(h) represent the eight sets of NURBS parameters used to define a scatterer shape. Further, each NURBS parameter set consists of two Q-Q plots of the control points ($C^x, C^y$) and one Q-Q plot of the corresponding weight ($w$).}
	\label{fig: qq_plot}
\end{figure}

The performance of the trained GRIDS-Net is evaluated on a set of test cases. The test dataset consists of 100 scatterers unknown to the trained network. Fig.~\ref{fig: remote_sensing_results} shows the visual comparison of the predicted and true scatterer shapes for nine randomly chosen scatterers from the test dataset. Fig.~\ref{fig: qq_plot} shows the Q-Q plots of the NURBS parameters on the complete test dataset. 

In Fig.~\ref{fig: remote_sensing_results}, we compare the predicted and true scatterer shapes and their corresponding $\mathbf{C}$. The high visual similarity between the predicted and true scatterer shapes provides a qualitative indication of the high quality predictions obtained by the trained GRIDS-Net. 
The high quality of these results can be directly attributed to its PBR module, which plays a significant role in enforcing the additional observational and learning biases that force the model to converge to a physically plausible design solution.

Fig.~\ref{fig: qq_plot}(a)-(g) show the Q-Q plots that provide an immediate indication of the overall prediction accuracy of the NURBS parameters ($\hat{\mathbf{C}}$ and $\hat{w}$) across all samples in the test dataset. In Q-Q plots, the closer the predictions are to the dotted straight lines, the higher the chances of predicted and true samples belonging to the same probability distribution. While the Q-Q plots show some differences in the probability distribution of predicted and true NURBS parameters, we observe significantly lower variation in the predicted control points ($\hat{\textbf{C}}_{0-7}$) compared to the predicted weights ($\hat{w}_{0-7}$). This observation is corroborated by the good prediction accuracy of $\hat{\textbf{C}}_{0-7}$ in Fig.~\ref{fig: remote_sensing_results}. However, the marked difference in predicted $\hat{w}_{0-7}$ is attributed to its negligible contribution to the determination of the scatterer shape. In other words, the scatterer shape is fairly insensitive to changes in $\hat{w}_{0-7}$. In summary, although a variation in predicted and true $\hat{\textbf{C}}_{0-7}$ and $\hat{w}_{0-7}$ exists, the qualitative assessments highlight a good match between the predicted and actual scatterer shapes. In addition, the results of the qualitative assessments on the test dataset also illustrate the outstanding generalization capability of the proposed inverse design approach.

\begin{figure}[h]
	\centering
	\includegraphics[width=1.0\linewidth]{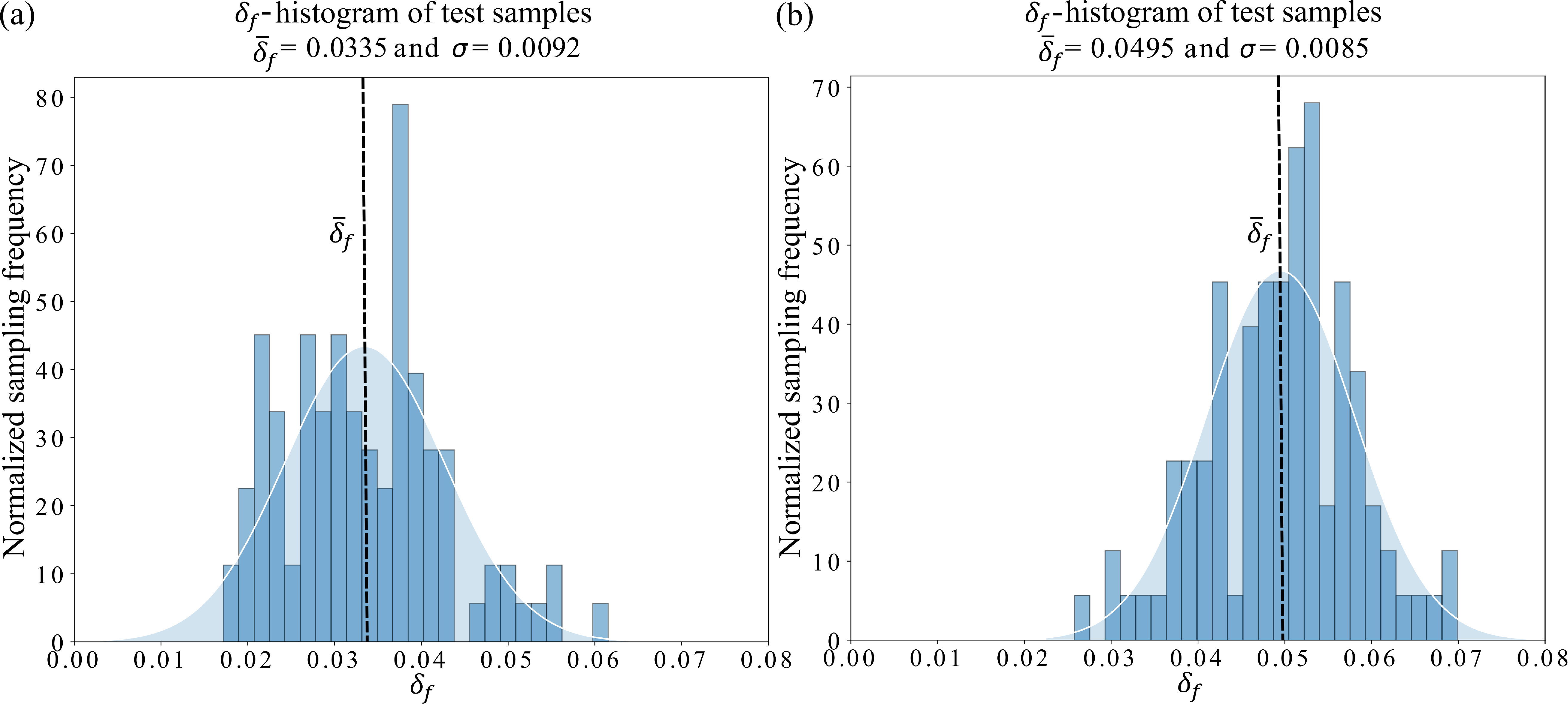}
	\caption{The $\delta_f$-histograms on the test dataset for the (a) GRIDS-Net predicted scatterer shapes and (b) modified GRIDS-Net (GRIDS-Net without the PBR module) predicted scatterer shapes. The solid white lines in (a) and (b) are the fitted normal distributions with mean $\bar{\delta}_f$ and standard deviation $\sigma$. The lower $\bar{\delta}_f$ value in (a) compared to (b) can be attributed to the physics-embedded learning in GRIDS-Net (using the PBR module), which regularizes the training process to predict accurate scatterer shapes. }
	\label{fig: DFD_plot}
\end{figure}

The $\delta_f$-histogram in Fig.~\ref{fig: DFD_plot}(a) plots the $\delta_f$ between the GRIDS-Net predicted and the true scatterer shapes for all the samples in the test dataset. In an ideal scenario, $\delta_f$=0 when the predicted and true shapes overlap, but due to errors associated with shape prediction $\delta_f > 0$ in all real scenarios. Hence, high shape prediction accuracy can be associated with lower $\delta_f$ values. A fitted normal distribution on the $\delta_f$-histogram (Fig.~\ref{fig: DFD_plot}(a)) evaluates a mean $\bar{\delta}_f=0.0335~m$ with standard deviation $\sigma=0.0092~m$. The small $\bar{\delta}_f$ value indicates a good match between the predicted and true shapes.
In summary, the qualitative (visual comparison, Q-Q plot) and quantitative ($\delta_f$-histogram) assessments, comparing the predicted and true scatterer shapes, clearly showcase the ability of the GRIDS-Net based inverse design approach to predict accurate scatterer shapes for remote sensing applications.  

\subsection{Inverse acoustic scatterer design: numerical results and  applications}
\label{ssec: material design}

In this section, the GRIDS-Net is used to solve the inverse problem of designing acoustic scatterers to mold the acoustic field in prescribed ways. In other words, the GRIDS-Net predicts the scatterer shapes that can produce predefined target acoustic fields.
Note that the accuracy in the target acoustic field can vary with the selected frequency of the incident acoustic wave. This condition is more conveniently restated in terms of the wavelength of the incident wave and the size of the scatterer via the parameter $kl_0^c$. When $kl_0^c<1$ (long wavelength or low frequency), the scattered wave is strongly affected by the cross-sectional area of the scatterer. At $kl_0^c \geq 1$ (short wavelength or high frequency), the geometric details of the scatterer surface start affecting the target acoustic field. Thus, at lower frequencies, a predicted scatterer with approximately the same cross-scattering area as the true scatterer can produce accurate target acoustic fields, but at higher frequencies, more accurate scatterer shape predictions are needed to capture the target response.
Additionally, this numerical study allows emphasizing the importance of physics-based regularization by developing a modified GRIDS-Net for inverse design. In a broad sense, a modified GRIDS-Net model is a GRIDS-Net model without the PBR module. We train the modified GRIDS-Net model with the same dataset as GRIDS-Net and compare the target acoustic fields produced by scatterer shapes predicted by GRIDS-Net and modified GRIDS-Net.   

Similar to  Fig.~\ref{fig: DFD_plot}(a), the $\delta_f$-histogram in Fig.~\ref{fig: DFD_plot}(b) is a quantitative measure of the similarity between the modified GRIDS-Net predicted and the true scatterer shapes across all the samples in the test dataset. A fitted normal distribution on the $\delta_f$-histogram (Fig.~\ref{fig: DFD_plot}(b)) evaluates a mean $\bar{\delta}_f=0.0495~m$ with standard deviation $\sigma=0.0085~m$. A direct comparison of $\bar{\delta}_f$ values from Fig.~\ref{fig: DFD_plot}(a) and (b) indicate a 32\% higher shape prediction accuracy of the GRIDS-Net in comparison to the modified GRIDS-Net model. This can be attributed to the presence of the PBR module in GRIDS-Net, which regularizes the learning process to predict scatterer shapes with higher accuracy.

\begin{figure}[h]
	\centering
	\includegraphics[width=1.0\linewidth]{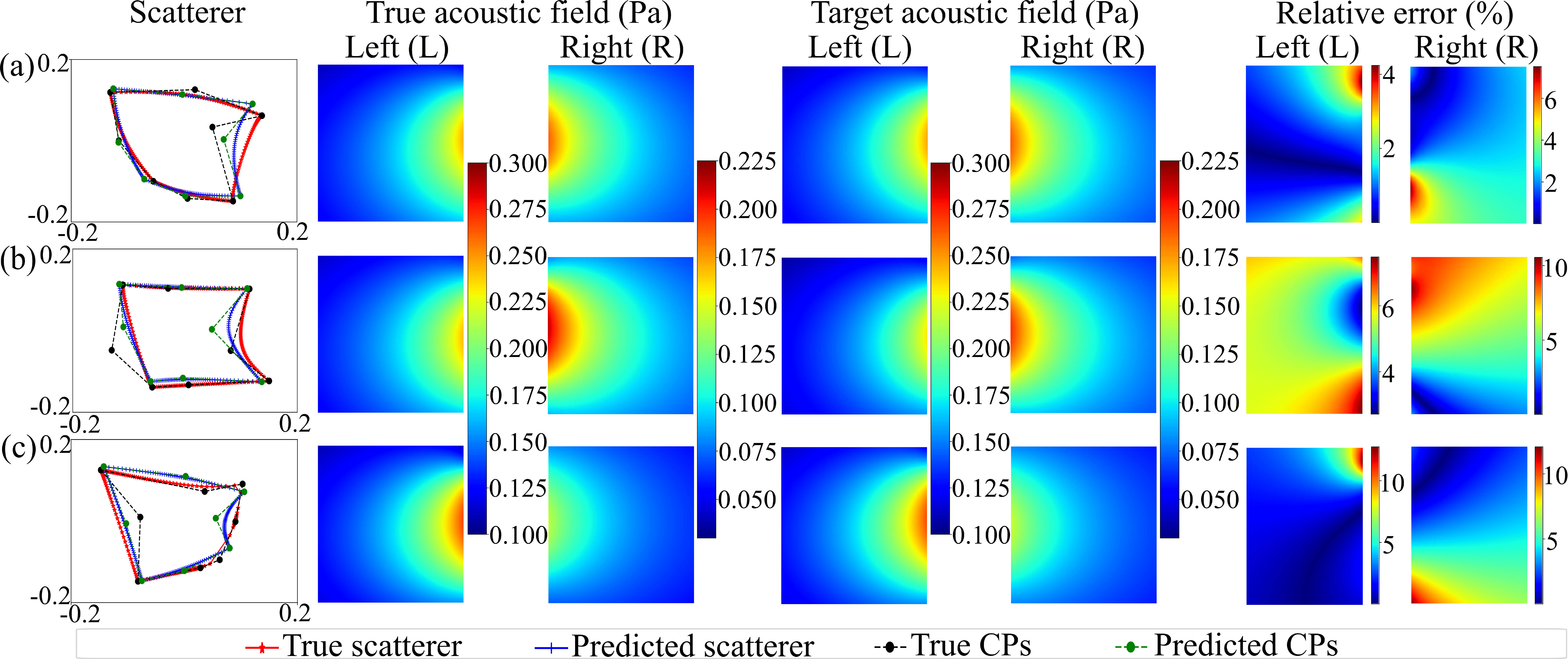}
	\caption{Amplitude maps of the true acoustic field, the target acoustic field, and the relative error \% between the true and target fields. (a)-(c) Three different scatterer shapes are considered and evaluated based on the acoustic field in the Left (L) and Right (R) target windows. The true acoustic field is the scattered pressure field due to the \textit{True scatterer}, while the target acoustic field is the scattered pressure field due to the \textit{Predicted scatterer}. The scattered acoustic responses were evaluated at an incident acoustic wave frequency $f=100~Hz$ ($kl_0^c < 1$) and at a unit value of the incident pressure.}
	\label{fig: material_design_results_100}
\end{figure}

\begin{figure}[h]
	\centering
	\includegraphics[width=1.0\linewidth]{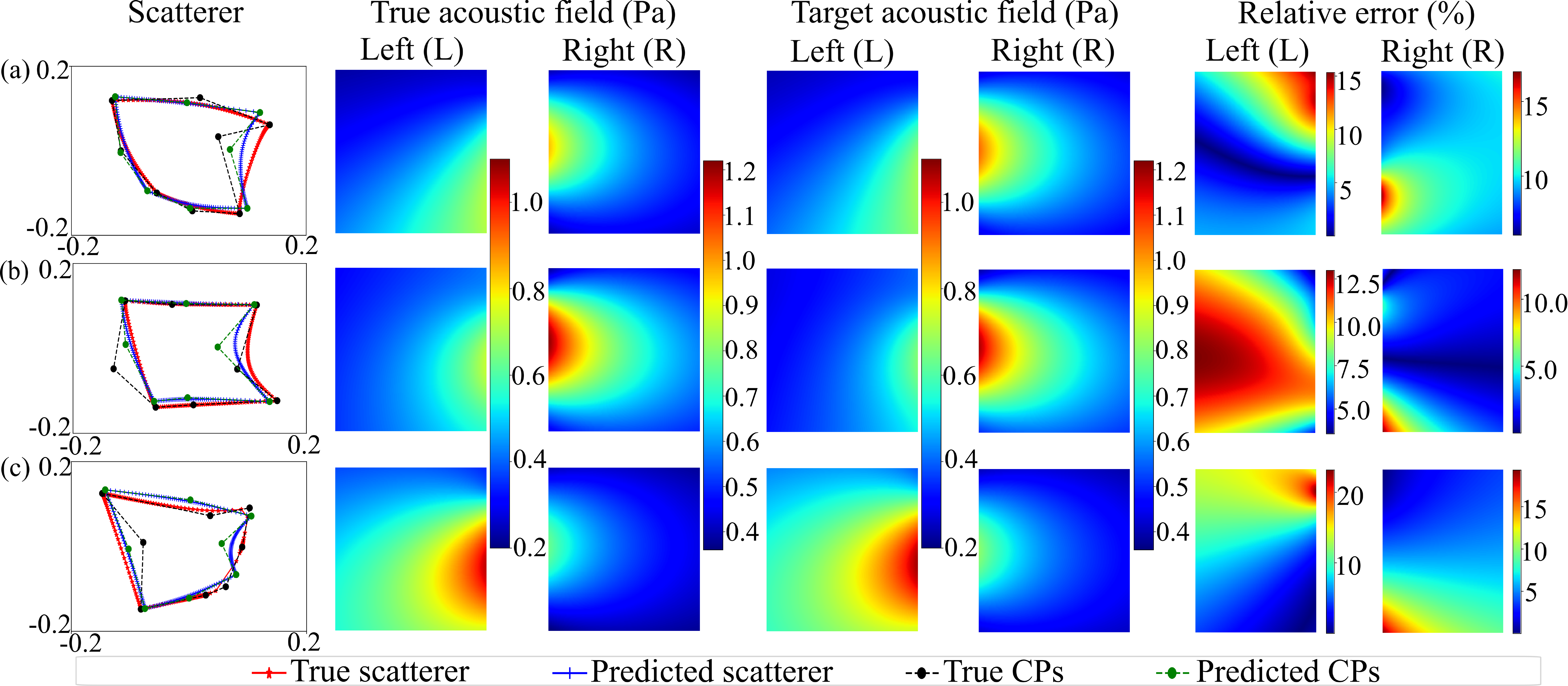}
	\caption{Amplitude maps of the true acoustic field, the target acoustic field, and the relative error \% between the true and target fields. (a)-(c) Three different scatterer shapes are considered and evaluated based on the acoustic field in the Left (L) and Right (R) target windows. The true acoustic field is the scattered pressure field due to the \textit{True scatterer}, while the target acoustic field is the scattered pressure field due to the \textit{Predicted scatterer}. The scattered acoustic responses were evaluated at an incident acoustic wave frequency $f=500~Hz$ ($kl_0^c \approx 1$) and at a unit value of the incident pressure.}
	\label{fig: material_design_results_500}
\end{figure}

Fig.~\ref{fig: material_design_results_100} compares the amplitudes of the target acoustic fields to the true acoustic fields and plots the corresponding relative error \% for three random scatterer samples extracted from the test dataset. This study compares the target and true acoustic fields evaluated on the Left (L) and Right (R) target windows as these two windows capture significant details of the scattered pressure response, as shown in Fig.~\ref{fig: surrogate_model}(a), for a planar incident wave travelling in the x-direction. The acoustic fields are evaluated at $f=100~Hz$ ($kl_0^c<1$). The mean relative error \% for the target fields in Fig.~\ref{fig: material_design_results_100}(a)-(c) are L: $0.92\%$, $5.37\%$, and $1.63\%$ and R: $2.40\%$, $5.47\%$, and $3.83\%$, respectively. Similarly, Fig.~\ref{fig: material_design_results_500} compares the target acoustic fields to the true acoustic fields for the same scatterer shapes at $f=500~Hz$ ($kl_0^c \approx 1$). The mean relative error \% for the target fields in Fig.~\ref{fig: material_design_results_500}(a)-(c) are L: $4.54\%$, $10.51\%$, and $8.12\%$ and R: $9.84\%$, $2.02\%$, and $5.89\%$, respectively. In summary, all the results show a mean relative error around or below $10\%$, hence illustrating the excellent ability of the GRIDS-Net model to identify scatterer shapes capable of capturing predefined target acoustic fields. The direct comparison of the same scatterer shapes produced by the modified GRIDS-Net indicates high prediction errors. This lower shape prediction accuracy of the modified GRIDS-Net manifests in the high target acoustic field error. Therefore, the target acoustic fields evaluated using the scatterer shape predicted by the modified GRIDS-Net accumulate higher errors than the target acoustic fields evaluated using GRIDS-Net predicted shapes.  Although the use of GRIDS-Net to predict a more accurate scatterer shape may not significantly enhance the target acoustic field accuracy at low frequencies, at higher frequencies ($kl_0^c \geq 1$) we expect the GRIDS-Net predicted scatterer shapes to more accurately capture the target acoustic fields than the modified GRIDS-Net predicted scatterers.

\section{Conclusions}
\label{ssec: Conclusion}

This study presented an end-to-end deep learning based inverse design approach, denominated neural network for the geometric regularization-based inverse design of shapes (GRIDS-Net), for the design of arbitrary-shaped 2D acoustic scatterers. Existing deep learning based approaches for inverse scatterer design either partially rely on traditional optimization schemes (thus, inheriting their limitations) or classify the scatterers into a predefined discrete set of shapes (hence, limiting the design space). This study addressed these limitations specifically by developing a regression-based end-to-end DNN capable of mapping the target pressure field to a continuous range of scatterer shapes. Although GRIDS-Net was explicitly applied to predict rigid scatterer shapes in acoustic scattering applications, the framework is general and could be extended to other classes of inverse design problems. GRIDS-Net addresses three major challenges associated with inverse shape design problems: the strong nonlinearity, the high dimensionality, and the ill-posedness. The integration of NURBS-based geometry parameterization within a convolutional autoencoder (CAE) architecture efficiently allows addressing the nonlinearity and the high dimensionality aspects of the inverse problem. In addition, geometric regularization via NURBS and physics-based regularization with BEM were integrated to address its ill-posed nature. The physics-based regularization was implemented by embedding GRIDS-Net with additional data that accounted for the physical behavior of the acoustic scattering problem. The inverse design approach was tested on two different classes of applications, including acoustic remote sensing and acoustic inverse material design. The validation and performance assessments were based on both qualitative and quantitative metrics. Overall, it was found that, for both classes of problems, the network predictions were in good agreement with ground truth solutions. Results emphasize that the integration of geometric regularization into DNNs is a key step to increase efficiency and accuracy in inverse design problems. While the numerical results highlight the importance of introducing geometric regularization via NURBS, they also emphasize the importance of physics-based regularization by directly comparing shape predictions obtained with and without the additional guidance of physics-based data. Remarkably, the proposed GRIDS-Net approach can predict any arbitrary shape within the spatial interpolation range of the training data, which means that the trained GRIDS-Net is capable of shape generalization. In addition, GRIDS-Net can easily adapt to predict inverse scatterer designs for arbitrary incident planar wave conditions as the integrated BEM algorithm facilitates the evaluation of additional data required for physics-based regularization. This is especially useful for the inverse design of scatterers, as any parameter other than the target pressure fields is considered an unknown of the inverse problem.

GRIDS-Net provides a solid foundation to develop advanced deep learning based inverse design approaches by integrating geometric regularization. Being predominantly a data-driven approach, which integrates a weakly informed physics model, GRIDS-Net can only interpolate (although continuously) between different scatterer shapes. It is reasonable to envision that the methodology could be further developed by integrating stronger physics-based constraints that could lead to scalable shape predictions with extrapolation capabilities.

\bigskip

\noindent \textbf{Competing interests}

The authors declare no competing interest.\\

\noindent \textbf{Acknowledgements}

This work was supported by the Laboratory Directed Research and Development program at Sandia National Laboratories, a multimission laboratory managed and operated by National Technology and Engineering Solutions of Sandia LLC, a wholly owned subsidiary of Honeywell International Inc. for the U.S. Department of Energy’s National Nuclear Security Administration under contract DE-NA0003525. The authors also acknowledge Mr. Mehdi Jokar for supporting the development of an early version of the BEM algorithm.  

\bibliographystyle{unsrt} 
\bibliography{Finaldraft}

\end{document}